# Characteristics and performance of the Multigap Resistive Plate Chambers of the EEE experiment

F. Coccetti[1, *] Abbrescia[2,3] C. Avanzini[1,4] L. Baldini[1,4,5] R. Baldini Ferroli[1,6]
G. Batignani[1,4,5] M. Battaglieri[1,7,8] S. Boi[1,9,10] E. Bossin[11] F. Carnesecchi[1,12,13]
C. Cicalò[1,10] L. Cifarelli[1,12,13] E. Coccia[1,14] A. Corvaglia[1,15] D. De Gruttola[16,17]
S. De Pasquale[16,17] F. Fabbri[1,6] D. Falchieri[13] L. Galante[1,18,19] M. Garbini[1,13] G. Gemme[8]
I. Gnesi[1,20] S. Grazzi[1] D. Hatzifotiadou[1,11,13] P. La Rocca[1,21,22] Z. Liu[23] L. Lombardo[24]
G. Mandaglio[1,21,25] G. Maron[26] M. N. Mazziotta[5] A. Mulliri[9,10] R. Nania[1,13] F. Noferini[1,13]
F. Nozzoli[27] F. Palmonari[1,12] M. Panareo[15,28] M. P. Panetta[1,15] R. Paoletti[6,29] M. Parvis[24]
C. Pellegrino[1,26] L. Perasso[1,8] O. Pinazza[1,13] C. Pinto[1,21,22] S. Pisano[1,6] F. Riggi[1,21,22]
G. Righini[1] C. Ripoli[16,17] M. Rizzi[5] G. Sartorelli[1,12,13] E. Scapparone[1,13] M. Schioppa[20,30]
A. Scribano[29] M. Selvi[1,13] G. Serri[1,9,10] S Squarcia[8,31] M. Taiuti[8,31] G. Terreni[1,4]
A. Trifirò[1,21,25] M. Trimarchi[1,21,25] C. Vistoli[26] L. Votano[14] M. C. S. Williams[1]
A. Zichichi[1,12,13] R. Zuyeuski[1] on behalf of EEE collaboration

[1]*Museo Storico della Fisica e Centro Studi e Ricerche "Enrico Fermi", Roma, Italy*
[2]*Dipartimento Interateneo di Fisica, Università di Bari, Bari, Italy*
[3]*INFN Sezione di Bari, Bari, Italy*
[4]*INFN Sezione di Pisa, Pisa, Italy*
[5]*Dipartimento di Fisica, Università di Pisa, Pisa, Italy*
[6]*INFN Gruppo Collegato di Cosenza, Laboratori Nazionali di Frascati (RM), Italy*
[7]*Thomas Jefferson National Accelerator Facility, Newport News, VA 23606, USA*
[8]*INFN Sezione di Genova, Genova, Italy*
[9]*Dipartimento di Fisica, Università di Cagliari, Cagliari, Italy*
[10]*INFN Sezione di Cagliari, Cagliari, Italy*
[11]*CERN, Geneva, Switzerland*
[12]*Dipartimento di Fisica, Università di Bologna, Bologna, Italy*
[13]*INFN Sezione di Bologna, Bologna, Italy*
[14]*Gran Sasso Science Institute, Italy*
[15]*INFN Sezione di Lecce, Lecce, Italy*
[16]*Dipartimento di Fisica, Università di Salerno, Salerno, Italy*
[17]*INFN Gruppo Collegato di Salerno, Salerno, Italy*
[18]*Dipartimento di Scienze Applicate e Tecnologia, Politecnico di Torino, Torino, Italy*
[19]*INFN Sezione di Torino, Torino, Italy*
[20]*INFN Gruppo Collegato di Cosenza, Laboratori Nazionali di Frascati (RM), Italy*
[21]*INFN Sezione di Catania, Catania, Italy*
[22]*Dipartimento di Fisica e Astronomia, Università di Catania, Catania, Italy*
[23]*ICSC World laboratory, Geneva, Switzerland*
[24]*Dipartimento di Elettronica e Telecomunicazioni, Politecnico di Torino, Torino, Italy*
[25]*Dipartimento di Scienze Matematiche e Informatiche, Scienze Fisiche e Scienze della Terra, Università di Messina, Messina, Italy*
[26]*INFN-CNAF, Bologna, Italy*
[27]*INFN Trento Institute for Fundamental Physics and Applications, Trento, Italy*
[28]*Dipartimento di Matematica e Fisica, Università del Salento, Lecce, Italy*

*Corresponding author.

[29]Dipartimento di Scienze Fisiche, della Terra e dell'Ambiente, Università di Siena, Siena, Italy
[30]Dipartimento di Fisica, Università della Calabria, Rende (CS), Italy
[31]Dipartimento di Fisica, Università di Genova, Genova, Italy

*E-mail*: `fabrizio.coccetti@cref.it`

ABSTRACT: The Extreme Energy Events (EEE) experiment, dedicated to the study of secondary cosmic rays, is arguably the largest detector system in the world implemented by Multigap Resistive Plate Chambers. The EEE network consists of 60 telescopes distributed over all the Italian territory; each telescope is made of three MRPCs and allows to reconstruct the trajectory of cosmic muons with high efficiency and optimal angular resolution. A distinctive feature of the EEE network is that almost all telescopes are housed in High Schools and managed by groups of students and teachers, who previously took care of their construction at CERN. This peculiarity is a big plus for the experiment, which combines the scientific relevance of its objectives with effective outreach activities. The unconventional location of the detectors, mainly in standard classrooms of school buildings, with heterogeneous maintenance conditions and without controlled temperature and dedicated power lines, is a unique test field to verify the robustness, the low aging characteristics and the long-lasting performance of MRPC technology for particle monitoring and timing. Finally, it is reported how the spatial resolution, efficiency, tracking capability and stability of these chambers behave in time.

**Contents**

## 1. Introduction

The Extreme Energy Events (EEE) experiment [1] is a project of the *Museo Storico della Fisica e Centro Studi e Ricerche "Enrico Fermi"* [2], in collaboration with the Istituto Nazionale di Fisica Nucleare (INFN), CERN and the Italian Ministry of Education, University and Research (MIUR). The EEE Project successfully combines educational objective with scientific results.

The EEE project is conceived to study cosmic rays and related phenomena, through a synchronous sparse network of 60 tracking detectors located throughout the Italian territory and at CERN. The network covers over 10 degrees of latitude and 11 degrees of longitude and is updated regularly, since a few new stations are built every year at CERN.

The EEE telescopes form a sparse network in which each location is at distances between 15 meters and several kilometers from another nearby. Each station, which is called an "EEE Telescope", consists of three MRPCs. It is a version, dedicated to cosmic rays, of the type of detector successfully used for Time Of Flight (TOF) in many high-energy physics experiments at colliders. The area covered by the EEE MRPCs, if placed close together, would be about 230 m$^2$, which makes EEE the largest system in the world that uses MRPCs. For comparison, the area occupied by the Alice TOF system is approximately 144 m$^2$.

A distinctive feature of the EEE network is that the telescopes are housed in High Schools and managed by groups of students and teachers, who previously took care of their construction at CERN. This peculiarity is a big plus for the experiment, which combines the scientific relevance of its objectives with effective outreach activities. The unconventional location of the detectors, mainly in standard classrooms of school buildings, with heterogeneous maintenance conditions and without controlled temperature and dedicated power lines, is a unique test field to verify the robustness, the low aging characteristics and the long-lasting performance of MRPC technology for particle monitoring and timing. Every year, new schools join the Project even without hosting a detector, contributing to the project by collaborating with neighbouring schools equipped with the telescopes; at the moment more than 100 High Schools are taking part in this project.

Data collection is centralized, raw data are transmitted from all EEE telescopes to the INFN-CNAF data center, where they are immediately reconstructed and stored. The current analyses concern more than 100 billion candidate muon reconstructed tracks.

The EEE experiment achieved scientific results in several fields related to Cosmic Ray understanding, for example: observation of Forbush decreases [3], study of muon decay into up-going events [4], detection of extensive air showers [5], search for long distance correlations between EAS [6], study of cosmic muon anisotropy at sub-TeV scale [7]. Now, the research of the EEE collaboration focuses also on: long time-scale structure stabilites, strategies to reduce the Global Warming impact and a detailed GEANT-GEMC simulation for the EEE telescopes. These latter topics are detailed in the proceedings of the RPC 2020 Conference.

## 2. MRPCs and EEE Telescopes

EEE telescopes use specially designed MRPCs to achieve: efficiency close to 100%, good time resolution, long term stability, good localization and synchronization capabilities combined with easy assembly procedures and low production costs.

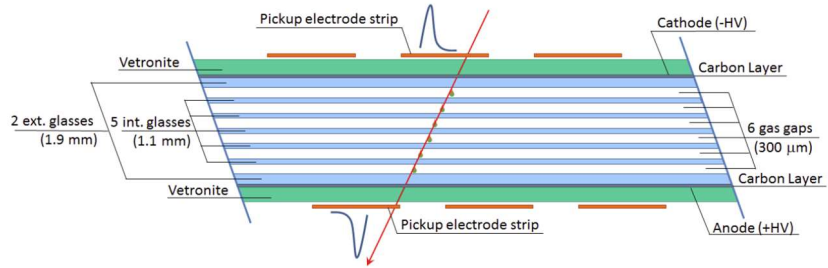

**Figure 1**. Cross section of one of the EEE Telescope's MRPC.

The detector structure (Figure 1) consists of 6 gas gaps obtained by stacking glass plates. High voltage is applied only to the external plates, leaving the internal ones floating. The cathode and anode are made of two glass plates measuring 160 cm by 85 cm, with a thickness of 1.9 mm. They are treated by students with resistive paint to build an electrode with a surface resistivity of 5–20 MΩ/□, which will be connected to high voltage. The total space between cathode and anode is divided into the six 300 μm spaces, through 5 intermediate glass plates each 158 cm wide, 82 cm high and 1.1 mm thick. Spacing between the internal glass plates consists of a weave made of fishing line. A sheet of Mylar is spread on both outer surfaces, on vetronite panels, both of size 175 cm by 86 cm. Twenty-four copper strips are placed here to collect the electrical signals induced by charged particles. The strips are 180 cm long, 2.5 cm wide and 0.7 cm apart from each other. These copper strips are applied on both sides of the stack of glass plates. Since the signal is given by the avalanches in the gas of all the spaces, this means that the signal that will be read is a differential signal of the cathode and anode strips. The signal of each strip is read by two Front-End Cards (FEA) positioned in the center of the narrow sides of the chamber. These FEA cards incorporate the ultrafast and low power NINO ASIC amplifier/discriminator specifically designed for MRPC operations. The 24 copper strips provide the two-dimensional information when a cosmic muon crosses the chamber: the y coordinate (short side) is determined by the strip on which the signal is induced, while the x coordinate (long side) is determined measuring the difference between the arrival time of the signal at the two ends of the strip. The arrival times of the signals are measured by two commercial TDCs made by CAEN, with 64 and 128 channels, and 100 ps bin. The mechanical stability of the entire structure is guaranteed by two rigid honeycomb composite panels measuring 180 cm by 90 cm. All these components are closed in a gas-tight aluminum box, with internal dimensions of 192 cm by 92 cm, and external of 220 cm by 110 cm.

High voltage connectors and gas inlets and outlets are located at the ends of the longer sides. High voltage to the chambers is provided by a set of DC/DC converters, with output voltage roughly a factor 2000 with respect to the driving low voltage. Stand-alone power supply units provide low voltage to the DC-DC converters. The core unit of the DC-DC converters are the EMCO Q-series, negative and positive, with a 10 kV full scale output. The HV stability declared by the manufacturer is ±10% at full load (50 µA). The working voltage of each DC-DC converter is from 8 to 9.5 kV, therefore the total HV applied on the chambers is in the 18 to 20 kV range.

The trigger logic consists of a six-fold coincidence of the OR signals from the FEA cards, corresponding to a triple coincidence of both ends of the chambers. Signals are combined in a VME custom made trigger module which was recently completely redesigned. The board was conceived to integrate in a double Eurocard board both a trigger logic and a GPS receiver for time stamping purpose. The GPS receiver is the ICM SMT 360™ unit, mounted on a carrier board in an open PCB assembly without enclosure, specifically designed to be integrated in host systems. This receiver can receive GNSS signals from GPS, GLONASS, Galileo, or Beidou satellite constellations. This new Trigger/GPS card is fully compatible with the DAQ system already in use in the experiment. Synchronization between telescopes is achieved via the GPS unit mentioned above, that provides the event time stamp so that all the telescopes are synchronized in the order of 10 ns. A collaboration with the Istituto Nazionale di Ricerca Metrologica (INRIM) has been recently set up, to further improve the performance of the synchronization across the EEE Network.

The chambers are filled with a mixture of 98% R134a ($C_2F_4H_2$) and 2% $SF_6$ gas, at a continuous flow and atmospheric pressure. The gas flow is provided by a commercial mixing system and fills the chambers in a daisy chain, with the exhaust connected to the outside. Until recently the flow was 2 l/h. To reduce the gas consumption in the detectors, a special procedure has been implemented to seal all the chambers, and to carry out a series of leak tests. At present, this operation was successfully completed for about half of the EEE telescopes. The ultimate goal is to reduce the flow of gas from 2 l/h to 1 l/h in all the EEE network. At the same time, a gas recirculation system is under study and will be implemented in the near future, to further reduce operating costs. Furthermore, all this adds up to studies on new gas mixtures to reduce the Global Warming impact that are in progress at CERN.

To reconstruct the tracks, a linear fit of the clusters found in the three chambers is performed and the corresponding $\chi^2$ is calculated. All possible cluster combinations are used and ordered by their $\chi^2$. The track candidates are defined by iteratively selecting the lowest $\chi^2$ and removing the corresponding clusters, continuing up to the point when the whole set of available clusters has been assigned to a track. Finally, a set of tracks with no hits in common is defined and transferred to the output file. For the measurements presented in this study, the track selection requires the cut $\chi^2 < 5$ and rejection of events with more than one track.

## 3. Performance

### 3.1 Time and Spatial resolution

Data taken in the last quarter of 2019 and first quarter of 2020 were used to compute time and spatial resolution. For this analysis, only a subset of the full network of telescopes was selected, specifically only those for which the gas flow was reduced from 2 l/h to 1 l/h through a MRPC chamber sealing procedure. These new results are compared with those obtained in previous years [8], with the flow of telescopes at 2 l/h, and are in agreement and do not show a drop in performance.

A comparison was performed between the hit information $s$ on the central chamber and the ones on the external chambers. The computed width of the distribution, $\Delta s$, is used to estimate the time resolution $\sigma_t$, the longitudinal spatial resolution $\sigma_x$ and the transverse spatial resolution $\sigma_y$ of the telescope. Time and spatial residuals used for the measurement of the time and spatial resolution are defined as $\Delta s=(s_{top}+s_{bot})/2-s_{mid}$, where $s_{top}$, $s_{mid}$, $s_{bot}$ are the time and spatial values for single or clustered hits and $s$ represents $t$, $x$, or $y$ respectively. More details can be found in [9]. Time resolution measurements are performed applying the Time Slewing correction. Figure 2 shows the comparison of the time resolution between the EEE telescopes taking data at 1 l/h and the same telescopes taking data al 2 l/h in 2017. In both cases, similar results are obtained, $\sigma_{t(1l/h)}=237\pm67$ps and $\sigma_{t(2l/h)}=238\pm40$ps, meaning that the sealing procedure of the detectors worked properly and has made it possible to reduce the flow of gas without affecting the time resolution.

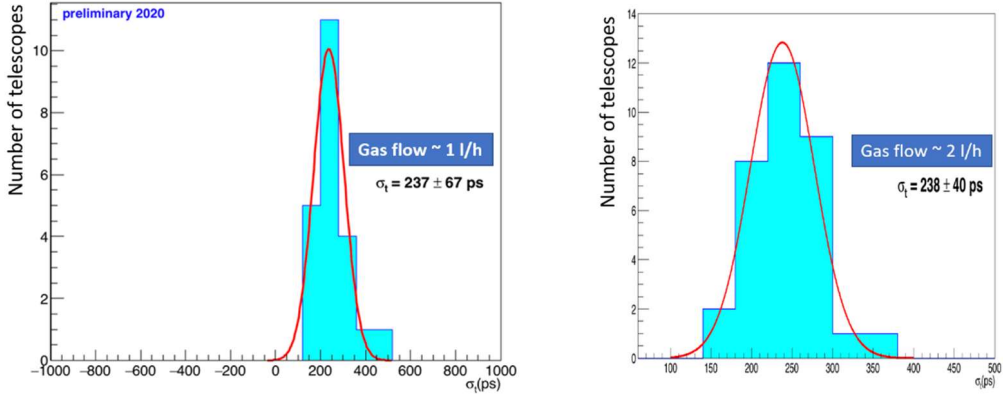

**Figure 2. Left:** Time resolution measured with data taken in the last quarter of 2019 and first quarter of 2020 for 22 telescopes, fed with gas flow of 1 l/h. The average time resolution is given by a gaussian fit. **Right:** Time resolution measured with data taken in RUN 3 (2017) for 33 telescopes, fed with gas flow of 2 l/h.

A completely similar analysis was made for spatial resolution. Figure 3 shows the longitudinal and spatial resolutions whose values are: $\sigma_{x(1l/h)}=1.4\pm0.1$ cm, for the telescopes with flow at 1 l/h and $\sigma_{x(2l/h)}=1.48\pm0.04$ cm, for gas flow at 2 l/h. The analogus plots, in Figure 4, refer to the transverse spatial resolution, where $\sigma_{y(1l/h)}=0.93\pm0.05$ cm and $\sigma_{y(2l/h)}=0.92\pm0.01$. Comparing the results of the spatial distributions of the detectors, it is seen that, even in the case of spatial resolution, the reduction of the gas flow did not cause a drop in performance.

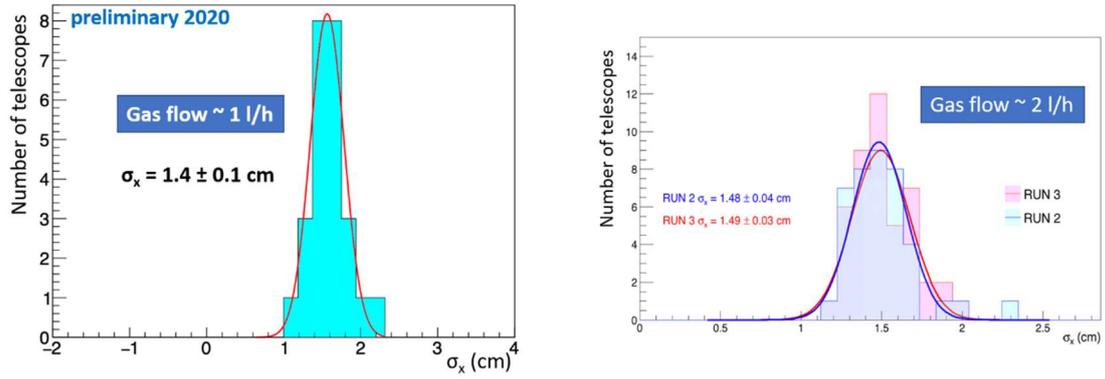

**Figure 3. Left:** Longitudinal spatial resolution, measured with data taken in the last quarter of 2019 and first quarter of 2020 for 25 telescopes, fed with gas flow of 1 l/h. The average time resolution is given by a gaussian fit. **Right:** Longitudinal spatial resolution measured with data taken in RUN 3 (2017) and RUN 2 (2016) for 46 telescopes, fed with gas flow of 2 l/h.

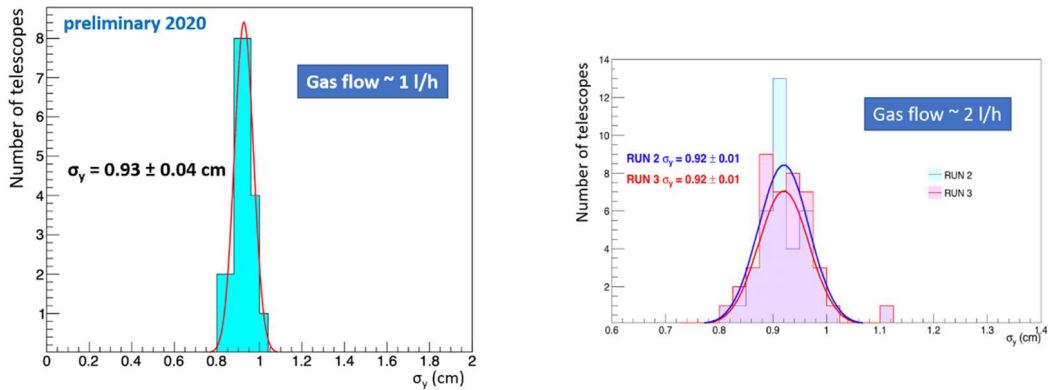

**Figure 4. Left:** Transverse spatial resolution, measured with data taken in the last quarter of 2019 and first quarter of 2020 for 25 telescopes, fed with gas flow of 1 l/h. The average time resolution is given by an RMS fit. **Right:** Transverse spatial resolution measured with data taken in RUN 3 (2017) and RUN 2 (2016) for 46 telescopes, fed with gas flow of 2 l/h.

### 3.2 Efficiency

The efficiency curves as a function of the applied voltage are measured both at CERN, immediately after the construction of the chambers, and after the installation of the telescopes in schools. In most cases these curves are obtained using scintillator detectors, used as an external trigger and with additional electronics. MRPC efficiency can also be measured during data taking, using a slightly modified version of the reconstruction code. In particular, the trigger logic is modified from the standard 3-chamber trigger to a double chamber coincidence, excluding the chamber under test from the trigger [8]. The two chambers in the trigger are also used for event selection and tracking. Once a track has been defined, the procedure checks whether there is a hit on the chamber under test within a distance of 7 cm from the calculated position. An HV scan of the chamber is performed, which collects around 150000 events per step. This method was applied to the middle chamber in the EEE telescopes, results are shown in Figure 5. The average efficiency of the telescope network is around 93%, compatible with EEE specs and with results of the beam test in [10]. An efficiency better than 90% is reached by 77% of the network.

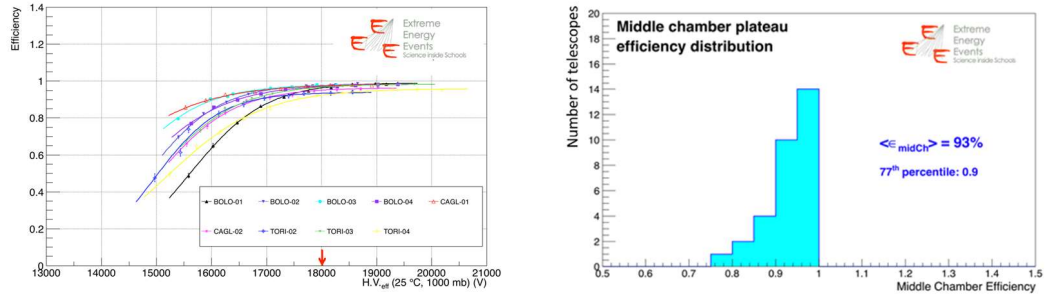

**Figure 5. Left:** Efficiency vs. applied HV (corrected for standard *p* and *T*) of the middle MRPC of 9 EEE telescopes. The red arrow indicates the usual value chosen as working point. **Right:** Distribution of the efficiency obtained at the plateau (corrected for standard p and T) of the middle MRPC for 31 EEE telescopes. An efficiency better than 90% is reached by 77% of the network.

### 3.3 Long term stability

In order to monitor the status of the telescopes, a special Data Quality Monitor has been set up, accessible to all the participants to the project. It is indeed difficult to achieve long-term stability for the EEE detectors, since many unexpected events can occur in schools. For this reason, the EEE collaboration continuously tries to improve the quality of the components that characterize the telescopes. The trigger card and GPS have recently been changed, while new DC-DC converters are under development. On the DQM website [11], several types of plots are available for all telescopes that report the most significant quantities in near real time, keeping all the history. In particular, it is constantly monitored: the average tracks $\chi^2$, the raw acquisition data, the multiplicity, the percentage of raw events where at least one-track candidate has been found, the average tracks Time of Flight between top and bottom chambers, the rate of events with at least one candidate track. No worrying effects of ageing in all these years have been noticed, although many of the telescopes are located in the most varied environments. This is proof of how reliable and long lasting the MRPC detectors are.

### 4. Conclusions

The Extreme Energy Events experiment, dedicated to the study of secondary cosmic rays, is arguably the largest detector system in the world implemented by Multigap Resistive Plate Chambers, which placed close together would cover an area of 230 m$^2$.
The performance of the telescopes is in accordance with the EEE experiment requirements in terms of efficiency (~93%), time resolution (238ps), spatial resolution (1.4 cm and 0.93 cm). In the coming years, more MRPCs will be built at CERN and more telescopes will be installed in schools; this will increase the coverage of the EEE network and the collected statistics, thus increasing the reach of the ongoing studies of this experiment.